\documentclass[
 reprint,
 amsmath,amssymb,
 aps,
 pra,
floatfix,
]{revtex4-2}

\usepackage{graphicx}
\usepackage{dcolumn}
\usepackage{bm}

\usepackage{mathtools}

\usepackage{amsmath}
\usepackage{graphicx}
\usepackage[colorlinks=true, allcolors=blue]{hyperref}
\usepackage{color,soul}
\usepackage{soul}

\begin{document}

    \title{Measurement and feedforward correction of the fast phase noise of lasers}

\author{T. Denecker\textsuperscript{1,2},
Y. T. Chew\textsuperscript{1},
O. Guillemant\textsuperscript{1},
G. Watanabe\textsuperscript{1,2},
T. Tomita\textsuperscript{1,2},
K. Ohmori\textsuperscript{1,2},
S. de Léséleuc\textsuperscript{1,3}
}

\affiliation{\textsuperscript{1}Institute for Molecular Science, National Institutes of Natural Sciences, Okazaki, Japan}
\affiliation{\textsuperscript{2}SOKENDAI (The Graduate University for Advanced Studies), Okazaki, Japan}
\affiliation{\textsuperscript{3}RIKEN Center for Quantum Computing (RQC), RIKEN, Wako, Japan}

\begin{abstract}
Lasers are the workhorse of quantum engineering in the atomic-molecular-optic community. However, phase noise of the laser, which can be especially large in popular semiconductor-based lasers, can limit gate fidelity. Here, we present a fully-fiberized instrument detecting and correcting the fast, sub-microsecond, phase fluctuations of lasers. We demonstrate a measurement noise floor of less than 0.1~Hz$^2$/Hz, and a noise suppression of more than 20~dB for Fourier frequencies in the 1 to 10 MHz region (reaching up to 30~dB at 3~MHz), where noise is critical for Rydberg-based quantum gates. 
Finally, we observe the improvement offered by this fast phase noise eater on a Raman transition driven by two such stabilized lasers. 
These measurement and correction techniques are important tools for high-fidelity manipulation of the excited electronic states of atoms and molecules.
\end{abstract}

\maketitle


\section{Introduction}
Over recent decades, quantum simulation and computation have advanced considerably across platforms, including trapped ions~\cite{ion_review_simu2021,ion_comp_review_2019}, superconducting qubits~\cite{sc_review_2020}, photonic qubits~\cite{pc_review_2009}, and cold atoms~\cite{Browaeys2020_review}. To ensure accurate outcomes of the machine, high-fidelity coherent control is crucial for each platform~\cite{Sung2021_sc_highfidelity,ballance2016high,evered2023highfidelity,Akerman2015}. 
While external factors often introduce decoherence, the control mechanisms themselves can also inadvertently contribute. 
Specifically, in neutral-atom or trapped-ion systems, where laser-driven dynamics of a valence electron play a central role, employing highly stable lasers is crucial.
Focusing on the problem of driving transition of neutral atoms to Rydberg states, the dominant decoherence sources~\cite{Graham_Saffman_2019, evered2023highfidelity, tsai2024} include the Doppler effect, spontaneous emissions \textit{via} the intermediate state used for the two-photon excitation, and laser intensity and phase noise \cite{SdL2018,Levine2018}—the latter being the focal point of this article.
Intuitively, the laser phase needs to be stable on a timescale of 0.1 to 1 microsecond during which the electron is manipulated (literally, shaken) by the laser field. A naive, but illustrative, requirement to suppress the error of a laser-driven quantum gate at the $10^{-3}$ level (0.1~\%) would be to ask for less than $ \phi_{\rm rms} = (2\pi) \times 10^{-3} = 6$~mrad of phase fluctuation in 1 microsecond. Assuming white frequency noise, this translates to a power spectral density (PSD) of phase noise $S_\phi(f)$ $\approx$ $\phi_{\rm rms}^2/f = 4 \times 10^{-11}$~rad$^2$/Hz at Fourier frequency $f = 1$~MHz, or equivalently a white frequency noise of $S_\nu(f) = f^2 S_\phi(f) = 40$~Hz$^2$/Hz. In fact, driven systems are more robust than this simple estimate, and recent analysis indicate that, at such level of PSD, error of manipulation are already suppress at the 0.01~\% level~\cite{Saffman_gatefidelity_2022,tsai2024}.  


While sources of electromagnetic radiation in the microwave domain can easily pass such phase noise requirements, this can be harder to achieve with lasers, often for fundamental reasons (the Schawlow-Townes limit). The best performers are bulk-cavity solid-state lasers (TiSapph, Nd:YAG, ...), with well below 1 Hz$^2$/Hz at MHz Fourier frequencies, thanks to their narrow-linewidth cavity design, thus offering highly-stable drives already appreciated by the Rydberg-atom community~\cite{Scholl2021, Graham2022}. But they pose other challenges in terms of wavelength coverage, as well as size and cost (SWaP-C). A popular alternative are semiconductor-based lasers, displaying a much wider coverage and high-level of integration. However, these lasers have a large quantum-limited frequency noise floor ($10^6$~Hz$^2$/Hz, for a basic Fabry-Perot design). While this can be mitigated by more advanced architecture, such as with extended-cavity diode lasers (ECDLs), their noise level ($10^2-10^4$~Hz$^2$/Hz) are still too large. Promisingly, novel designs that recently reached commercialization --- based on vertical-external-cavity surface-emitting laser (VECSEL)~\cite{Guina2017,Hastie2023}, or self-injection-locked (SIL) laser diodes~\cite{Maleki2015,zhang2024} ---, could pass the requirement. Easily measuring the frequency noise of such lasers is a first motivation of this work. 

Alternatively, the phase noise of a laser can be suppressed externally. A first solution is to pass the laser through a passive filter realized with a high-finesse Fabry-Perot cavity\cite{Hald2005,Nazarova2008}, which demonstrated great improvements~\cite{Levine2018}. A drawback of this approach is the limited output power and tunability. Another approach is to actively correct the laser's phase. There, feedback correction on the laser diode current or cavity length is commonly employed, and is hugely successful for suppressing slower technical noise from acoustic fluctuations or thermal drifts. However, this fails at suppressing the fast, sub-microsecond, phase fluctuation of interest here. The finite travel time of information around the feedback loop sets a maximum correction bandwidth of a few MHz~\cite{Schoof2001, Legouet2009,Appel2009,Marangoni2015,Endo2018,Birkl2022}. This limitation can be evaded if the correction is not fed back to the laser, but rather fed forward to a phase actuator~\cite{Hashemi2009,Hashemi2010,Lintz2017,Covey2022,Tey2023}. Correcting the fast phase noise with a feedforward approach is the second aspect of this work.

\section{OVERVIEW}
\label{sec:overview}

\begin{figure}[!ht]
    \includegraphics[width=0.95\columnwidth]{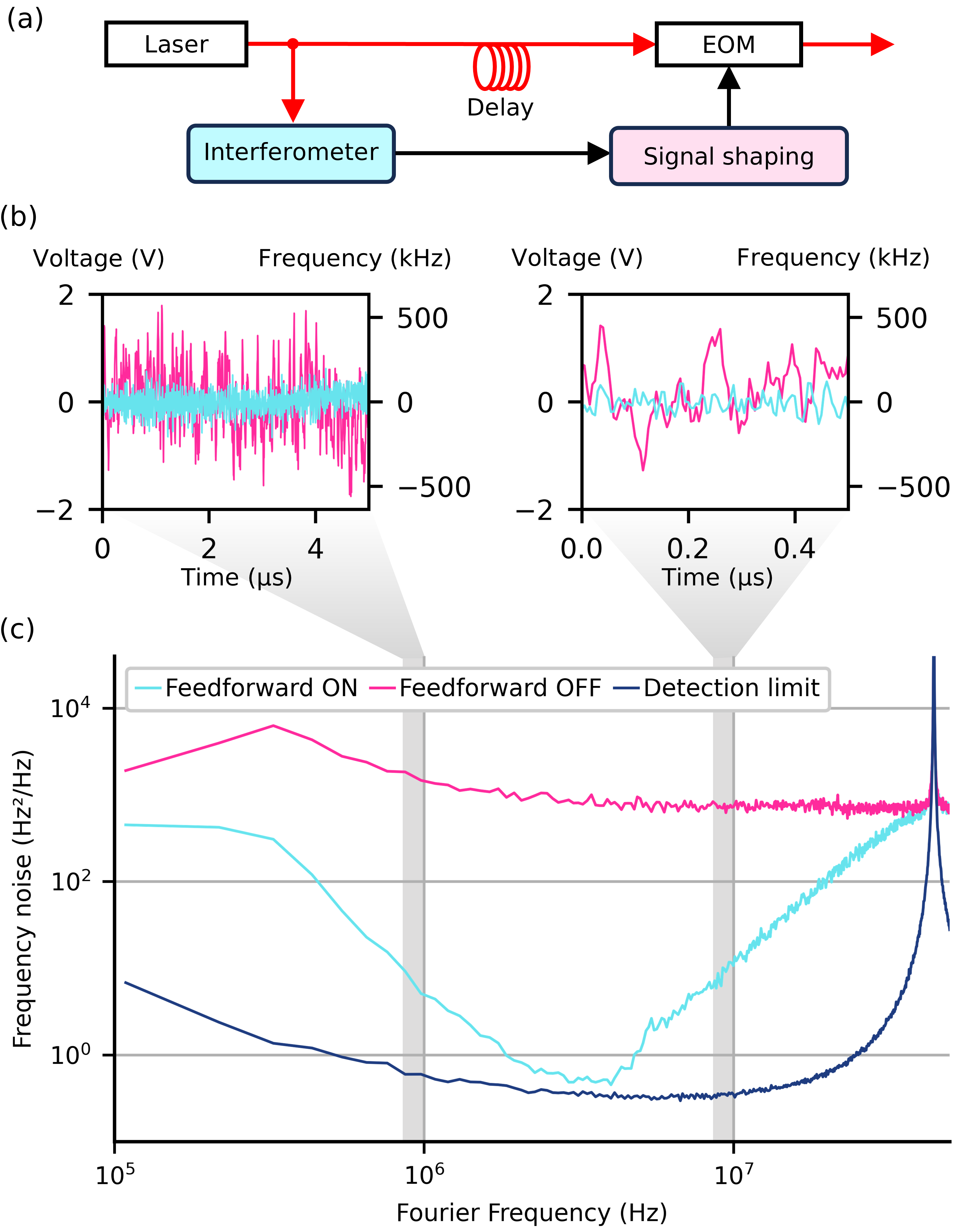}
    \caption{Measurement and correction of fast phase noise. (a) Schematic of the phase noise eater. (b) Time traces of the measured frequency noise with (in blue) and without (in pink) the feedforward correction. (c) Frequency noise PSD with and without the correction. The dark blue curve is the detection noise floor. A cancellation of more than 30~dB is achieved at a Fourier frequency of 3 MHz, where the noise is suppressed down to 0.5~Hz$^2$/Hz.}
    \label{fig:feedforward scheme}
\end{figure}

We first gathered in this section the main results of our work, with the more technical details of the phase noise measurement and correction covered later. Figure~\ref{fig:feedforward scheme} shows a simplified schematic of the fast phase noise eater, as well as its performance in the time and frequency domain. Following Fig.~\ref{fig:feedforward scheme}(a), a fraction of the laser light is sent to an interferometer, while the rest goes to a fiber delay line and then to a fast electro-optic modulator that will correct the phase. The interferometer converts the instantaneous frequency noise into a voltage, which is then reshaped to reproduce the phase noise and finally applied to the EOM. With precise adjustments of the feedforward signal, a reduction of the phase noise by up to a factor of 100 (40 dB in PSD) is conceivable (but challenging). All details are given in later sections. 

We now describe the results in the time domain shown in Fig.~\ref{fig:feedforward scheme}(b). We use a second interferometer, placed after the phase noise eater, to detect the frequency noise with (blue traces) and without (pink) activating the feedforward path. The output voltage, left vertical axis, is converted into frequency deviation, right vertical axis, by using the interferometer calibrated DC sensitivity  $s_0 = 3.1$~V/MHz. The left (right) panel shows the interferometer raw output, observed with an oscilloscope set on a microsecond (sub-microsecond) timescale. Clearly, the noise has been largely suppressed indicating good performance of the technique. 
The limits of the phase noise eater can also be distinguished from small noisy features remaining on very fast scales of tens of nanoseconds. The rms deviation is decreased from 200 kHz to 60 kHz, dominated by the faster noise.

For a more quantitative description, we investigate the correction in the Fourier domain with a spectrum analyzer (SA), whose output is shown in Fig.~\ref{fig:feedforward scheme}(c). Similar to the time domain, we convert the PSD measured by the SA into a frequency-noise PSD from the frequency-dependent sensitivity  of the interferometer (detailed later). The laser frequency noise with (without) the active correction is shown in blue (pink). The detection limit is also plotted in dark blue and obtained when no light is injected in the interferometer. The peak at 50 MHz originates from the null sensitivity of the interferometer. The unstabilized output has a white frequency noise of $S_\nu =700$~Hz$^2$/Hz, attributed to the fundamental quantum noise of this laser source. By activating the feedforward correction, the frequency noise is strongly suppressed in the 1 to 10 MHz region. The peak suppression is obtained at 3 to 4 MHz, where the PSD is decreased down to 0.5~Hz$^2$/Hz, which is $31.5$~dB below the initial noise level. The noise is remains below 10~Hz$^2$/Hz over the entire 1 to 10 MHz region, with the correction decreasing outside this range. The performance of the noise eater fully agrees with its model, presented later. Its peak performance lies in the region of maximal sensitivity to frequency noise of fast Rydberg quantum gates~\cite{tsai2024}, whose sensitivity drops quickly above the Rabi frequency of the laser drive (typically 3 - 5 MHz). At lower Fourier frequencies ($<1$~MHz), usual feed-back stabilization techniques are amply sufficient and it is thus acceptable for the feedforward correction to not be operative in this region.  


The rest of this paper gives details on the results that have been briefly summarized above.
In Section~\ref{sec:measurement}, we elaborate on the measurement of the phase/frequency noise with a delayed Mach-Zehnder interferometer (MZI), and present frequency noise measurements from a collection of lasers found in the laboratory. Then, in Section~\ref{sec:correction}, we detail the feedforward scheme, perform a quantitative comparison to its model, and discuss its performance relative to other reported results in the literature. Finally, in Section~\ref{sec:atom}, we shine the phase noise-canceled laser to atoms and observe an improvement on a Ramsey interferometry signal. 



\section{Phase noise measurement}
\label{sec:measurement}

In this section, we first discuss how to characterize phase noise and what method are used to measure it. We then focus on the realization of our fully fiberized delayed MZI to obtain the phase noise spectrum. Finally we present a collection of phase/frequency noise measurements for a variety of laser found in the laboratory. 

\subsection{Phase noise characterization}

A laser field with phase noise is described as:
$$E(t) = E_0 \cos{[2\pi \nu_0 t + \phi(t)]},$$ 
with $\phi(t)$ a fluctuating phase, and $\nu_0$ the (average) laser frequency. Alternatively, one can also consider the instantaneous laser frequency deviation $\nu(t) = (1/2\pi){\rm d}\phi/{\rm d}t$. We are interested in the variation of these quantities on timescale of 0.1 to 1 microsecond, or equivalently in the Fourier domain, around Fourier frequencies $f$ of 1 to 10 MHz, where the sensitivity of a Rydberg quantum gate to laser frequency noise is largest~\cite{tsai2024}. The adequate tool to quantify the noise of $\phi$ and $\nu$ is the PSD: $S_\phi$ (in rad$^2$/Hz) for the phase noise, or $S_\nu = f^2 S_\phi$ (in Hz$^2$/Hz) for the frequency noise. The PSD represents the variance of the signal contained in a unit frequency band of 1 Hz centered around $f$. Throughout this work, we will rather work with $S_\nu$ and also note that we use single-sided PSD (defined for positive Fourier frequencies only). The PSD can be input directly into a quantum gate error model based on linear response formalism~\cite{tsai2024}, or used to reconstruct random realizations of phase noise feeding numerical simulations~\cite{SdL2018,Saffman_gatefidelity_2022}.

\begin{figure}[!htbp]
    \centering
    \includegraphics[width=\columnwidth]{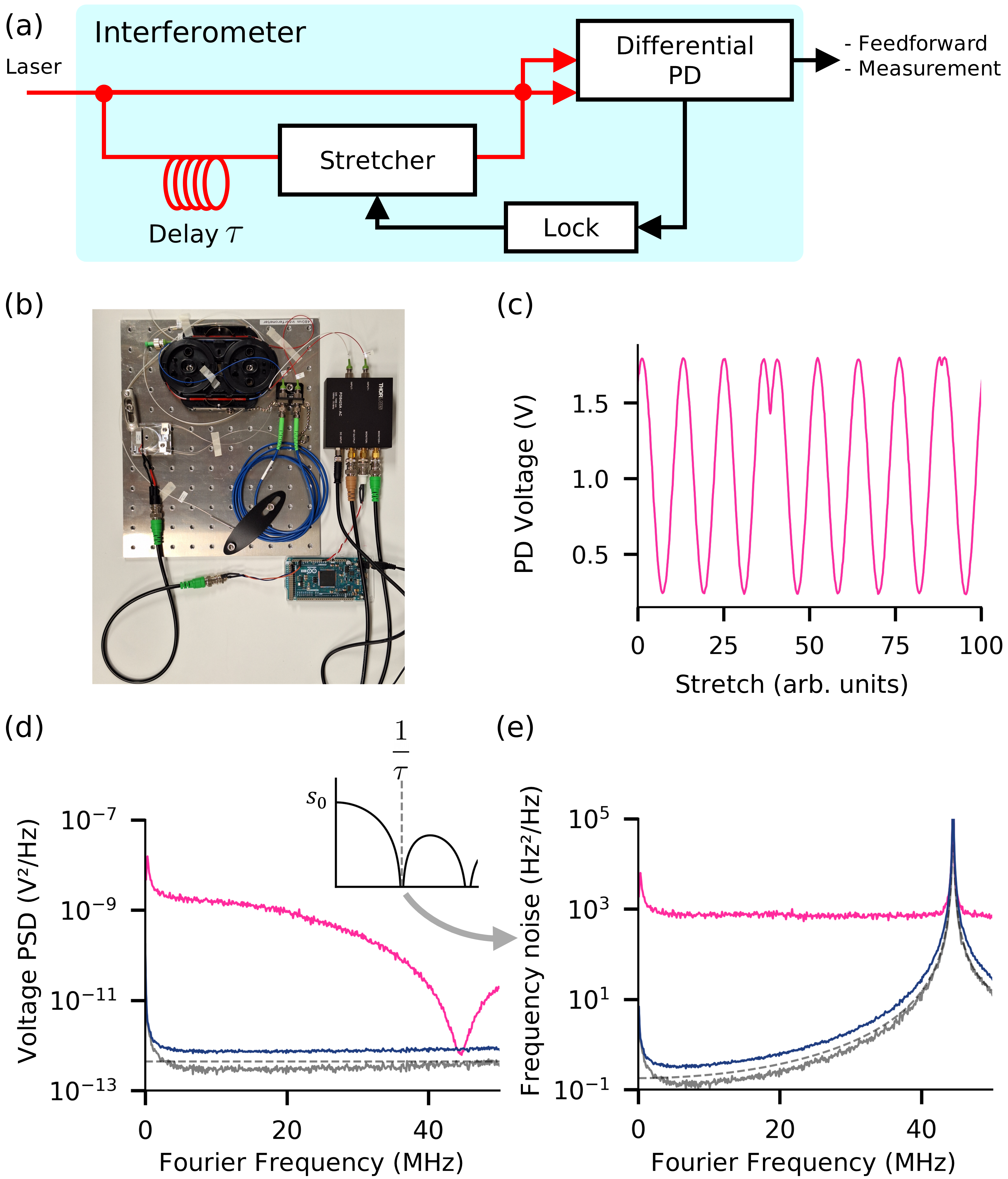}
    \caption{A delayed MZI as a fast phase/frequency noise discriminator. (a) Schematic and (b) picture of the realization with fiberized components. (c) Photo-diode (PD) output when scanning the stretcher. The peak-to-peak voltage $V_{\rm pp}$ is recorded to calibrate the interferometer sensitivity, and the interferometer is then locked at mid-point. 
    (d) The detector output is measured with a spectrum analyzer to give the PSD of voltage noise $S$ (in V$^2$/Hz) as a function of the Fourier frequency $f$. The curves in the semi-log plot correspond to the noise obtained when measuring an ECDL laser (pink), the electronic noise floor set by the transimpedance amplifier (gray, solid line), the photon shot noise (gray, dashed line) and the total detection noise floor (dark blue).
    Inset: MZI sensitivity $s$ to frequency-noise as a function of the Fourier frequency. The expected nulls of sensitivity at multiples of $f = 1/\tau$, clearly observed in (d), are used to precisely calibrate the MZI delay $\tau = 22.4$~ns. (e) Frequency noise PSD obtained after converting the spectrum analyzer output $S$ into a frequency-noise PSD $S_\nu = S/s^2$ using the calibrated MZI sensitivity.}
    \label{fig:MZI_discriminator}
\end{figure}

We now recall how the phase/frequency can be measured. For the microsecond timescale of interest, a rather simple delayed MZI is perfectly adequate. Figure~\ref{fig:MZI_discriminator}(a) shows the schematic, where the laser is first split in two arms, with one of them longer by a few meters (i.e., tens of nanoseconds), and finally recombined with their interference recorded onto a balanced photo-detector giving an electronic voltage $V(t)$. The output signal, obtained as a function of the phase difference between the two paths, is:
\begin{align}
\label{eq:MZI}
V(t) &= \frac{1}{2} V_{\rm pp} \cos[2\pi \nu_0 \tau + \phi(t) - \phi(t-\tau)] \nonumber \\  
&\simeq \frac{1}{2} V_{\rm pp} [\phi(t) - \phi(t-\tau)]
\end{align}
with $V_{\rm pp}$ the peak-to-peak voltage of the fringe,
and $\tau$ the delay between the two arms of the interferometer. The approximation on the second line is obtained when the MZI is operated at its quadrature point, giving maximal sensitivity  to frequency noise. 
For phase fluctuation slower than the delay $\tau$, we can approximate:
\begin{equation}
\label{eq:freqapprox}
\nu(t) = \frac{1}{2\pi} \frac{d\phi}{dt} \approx \frac{1}{2\pi} \frac{\phi(t) - \phi(t-\tau)}{\tau}
\end{equation}
to obtain the frequency-to-voltage conversion: 
\begin{equation}
\label{eq:DC_sensibility}
 V(t) \simeq (V_{\rm pp}  \pi \tau) \times \nu(t) = s_0 \times \nu(t)
\end{equation}
where we introduced the (DC) sensitivity  $s_0$.
The sensitivity  can be directly and accurately calibrated by measuring the peak-to-peak voltage $V_{\rm pp}$ of the MZI fringes on an oscilloscope, and extracting the delay $\tau$ from the zeroes of the MZI sensitivity  on a spectrum analyzer, see following discussion and Fig.~\ref{fig:MZI_discriminator}(c,d). Typically, we reach a sensitivity  of $\sim 3$~V/MHz depending on the choice of photo-detector (e.g., with/without a transimpedance amplifier). 

Equation~\ref{eq:freqapprox} is valid only for phase (or frequency) fluctuations much slower than the timescale set by the MZI delay $\tau$. Instead, we can derive an exact result by working in the Fourier domain. Denoting by $\tilde{V}$ the Fourier components of $V$ (and similarly for other quantities), we obtain:
\begin{align}
\label{eq:MZI_sensibility_derivation}
\tilde{V} &= \frac{V_{\rm pp}}{2} (1 - e^{- i 2\pi f \tau})\tilde{\phi}  \nonumber \\
&= s_0 \frac{\sin(\pi f \tau)}{\pi \tau} i e^{-i \pi f \tau} \tilde{\phi} \nonumber \\
&= s_0 \frac{\sin(\pi f \tau)}{\pi f \tau} e^{-i \pi f \tau} \tilde{\nu} 
\end{align}
leading to a frequency-dependent sensitivity :
\begin{equation}
\label{eq:MZI_sensibility}
s(f) = s_0 \frac{\sin(\pi f \tau)}{\pi f \tau} e^{-i \pi f \tau}
\end{equation}

We see from this equation, also shown in Fig.~\ref{fig:MZI_discriminator}(d), that the MZI display zeroes of sensitivity  at multiple of $1/\tau$, as well as an overall $1/f$-drop beyond the first zero. We typically use delays $\tau < 50$~ns, such that our region of interest ($f = 1 - 10$~MHz) is well within the first zero, which is especially important for the feedforward correction. However, for phase noise measurement, we can also extract information beyond $1/\tau$ (at the cost of a decreasing sensitivity).
We note that the choice of delay $\tau$ is a compromise between sensitivity  (increasing with $\tau$) and measurement bandwidth (decreasing with $\tau$).  

Before moving to the experimental realization, we shortly discuss how this frequency discriminator based on a short-delay MZI compares to other approaches. First, compared to delayed self-heterodyning with kilometer-long delays, often used to extract laser linewidth, the experimental setup is much simpler (to the exception that the interferometer needs to be actively stabilized to its quadrature point). Secondly, we point out that since we are only interested in information at high Fourier frequencies, there is no concern of technical noise (on the detectors, or on the fibers). Finally, if the lasers are already stabilized on a narrow linewidth cavity (as often the case to accurately set the average frequency $\nu_0$), it is possible to extract similar information from the Pound-Drever-Hall error signal~\cite{SdL2018}, or by beating the laser with the filtered output of the cavity~\cite{CoveyPRA2022, Tey2023}. 

\subsection{Fiber MZI realization}



We assembled a fiberized MZI from individual polarization-maintaining (PM) components, see Fig.~\ref{fig:MZI_discriminator}(b). 
The light is first split by a 50/50 PM coupler (Thorlabs, PN780R5F2) into a long and short arm. In the long arm, we place a delay line and a slow phase actuator (a home-made fiber stretcher with resonance beyond kHz). 
The two arms are then recombined on a second 50/50 coupler.  
We obtain interference fringes at the two outputs of the MZI, as shown in Fig.~\ref{fig:MZI_discriminator}(c), with visibility of typically 90~\% which are stable over time. The residual 10 \% loss of contrast is caused by imperfections of the components causing a slight imbalance of power and polarization between the two arms, but it does not affect the performance of the MZI as a frequency discriminator (only slightly decreases the sensitivity ). 
The MZI acts as a frequency discriminator when the two arms are combined with a phase difference of $\pi/2$, i.e., when the interferometer is at its quadrature point at the middle of the fringe. 

To lock the interferometer, we implemented a slow, sub-kHz, feedback loop with an Arduino DUE board monitoring the MZI output and controlling the fiber stretcher. We put a $\rm kHz$ low-pass filter (not pictured in Fig. \ref{fig:MZI_discriminator}) after the Arduino to prevent its electronic noise from exciting the resonance frequency of the fiber stretcher.

The two outputs of the last coupler are detected with photo-diodes. A balanced detection, where the two photo-currents from each output are subtracted, is preferred as it rejects (in first-order) the intensity noise from the laser while doubling the sensitivity to frequency noise. The photo-current can either be read directly on a spectrum analyzer, which allows the lowest measurement noise floor, or input to a transimpedance amplifier. We used the second approach (with the Thorlabs PDB425A) to drive the feedforward path, at the cost of a slight increase of noise floor (see next section). The amplifier bandwidth of 100~MHz is well beyond the MZI first zero of sensitivity.

\subsection{Laser frequency noise analysis}

We now turn to the analysis of the signal from the MZI containing information of the frequency/phase noise of the laser. We first discuss the calibration of the MZI sensitivity and then discuss the detection noise floor. 
\\
\paragraph{Calibration}
The full fringes are observed by scanning the fiber stretcher, see Fig.~\ref{fig:MZI_discriminator}(c), which allows to extract the peak-to-peak voltage $V_{\rm pp}$. Note that we show here the low-bandwidth amplified output of a single photo-diode, while for the spectral measurement we use the balanced signal from the two photo-diodes which has higher bandwidth and is further amplified.
Then, the interferometer is locked at mid-fringe, the quadrature point, allowing to observe frequency noise as voltage fluctuations shown in Fig.~\ref{fig:feedforward scheme}(b).

We then read-out the detector output with a spectrum analyzer to obtain the PSD of voltage $S$ (in V$^2$/Hz). The zeros of sensitivity  of the MZI appears strikingly at multiple of $1/\tau$~MHz , allowing to precisely calibrate the MZI delay $\tau = 22.4$~ns. Together with the measured peak-to-peak voltage $V_{\rm pp}$, this completes the characterization of the MZI sensitivity  $s_0 = V_{\rm pp} \pi \tau$. For the signals shown in Fig.~\ref{fig:MZI_discriminator}, we have $s_0 = 1.57$~V/MHz. This calibration procedure is simple and accurate, and is repeated for each interferometers or lasers used in this work. With this sensitivity, we can now convert $S$ (V$^2$/Hz) into the PSD of frequency noise $S_{\nu} = R/s^2 \times S$ (Hz$^2$/Hz). We note that for a careful quantitative measurement, one has to set the spectrum analyzer detection mode to average rms power (and not the default peak mode, or logarithmic average~\cite{rauscher_fundamentals_2016}), and to remember that a spectrum analyzer gives single-sided PSD in case one wants to convert it into two-sided PSD. 
\\
\paragraph{Noise floor}
When using a biased photo-detector for measurement, the electronic noise floor is first fundamentally set by the thermal Johnson noise of the $R = 50 \, \Omega$ impedance whose PSD is $S = 4k_BTR = -180$~dBm/Hz ($0.2$~nV/$\sqrt{\rm Hz}$) at room temperature. In practice, this noise source is overwhelmed by the spectrum analyzer electronic noise floor, typically at $S \simeq -160$~dBm/Hz (2~nV/$\sqrt{\rm Hz}$). This converts into a frequency noise floor of $S_\nu \simeq 5 \times 10^{-3}$~Hz$^2$/Hz. 
When using the transimpedance amplifier, the electronic noise increases to typically $0.1$~Hz$^2$/Hz, as shown with the gray solid line in Fig.~\ref{fig:MZI_discriminator}(d, e). In addition to the electronic noise, we can be further limited by the photon shot noise.
We evaluate it to $-110$~dBm/Hz for 200~$\mu$W of power onto the photo-detectors, leading to an equivalent frequency noise floor of $0.2~\rm Hz^2/Hz$, represented by the gray dashed line. Together, the electronic and photon shot noise add up to a detection limit represented in dark blue, giving 32 dB of signal-to-noise ratio.


In the Fourier frequency range of 0.1 to 10 MHz, we reached this noise floor.
At lower frequencies, we however observed an increase of noise. 
This is explained by the remaining electronic noise of the Arduino DUE after the $\rm kHz$ low-pass filter and the sensitivity of optical fibers to external perturbations. 
We conclude that this simple setup is perfectly adequate for measuring fast phase noise on the microsecond timescale and faster, but not for longer timescale.  



\subsection{Gallery of laser frequency noise}

\begin{figure}[t]
    \centering
    \includegraphics[width=\columnwidth]{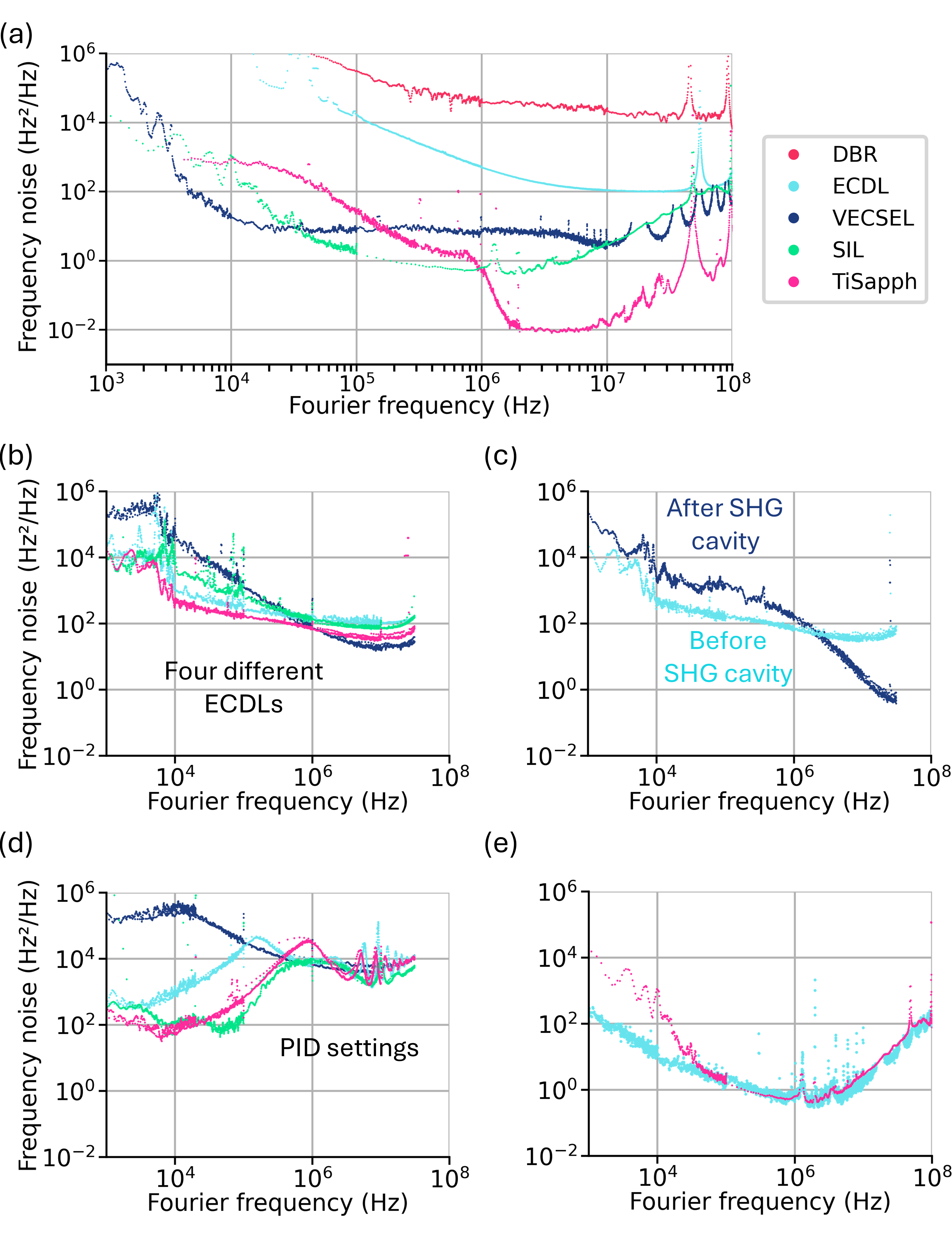}
    \caption{Gallery of laser frequency noise PSD. (a) Five lasers of different types: semiconductor-based DBR, ECDL, VECSEL, SIL, and solid-state-based TiSapph. (b) Four different ECDLs from the same manufacturer. (c) Frequency noise before and after frequency doubling with SHG in an enhancement cavity. (d) Noise reduction using a feedback loop with different settings of the PID parameters. (e) Noise measured with the home-made MZI (pink) and a commercial phase noise analyzer (blue).}
    \label{fig:noisecollection}
\end{figure}

We now show a collection of frequency noise for lasers found in the laboratory and under various conditions. 
We start by comparing the noise of five different types of lasers on Fig.~\ref{fig:noisecollection}(a): a DBR (distributed Bragg reflector) semiconductor diode laser from Vescent/Photodigm, an ECDL from Toptica, a VECSEL from Vexlum, a SIL laser from OEWaves, and finally a solid-state TiSapph laser from Sirah. 

As expected, the fast frequency noise of these lasers drastically depends on their architecture, with the DBR laser reaching a white frequency noise floor of 10 000 Hz$^2$/Hz while the solid-state laser noise floor is below the sensitivity  of 0.01~Hz$^2$/Hz of our detection scheme. 
The ECDL is around 100~Hz$^2$/Hz, limited as the DBR by the fundamental quantum noise of lasers given by the Schawlow-Townes limit: $S_\nu^{\rm qu} = h \nu_0 \nu_c^2 /P \times (1 + \alpha^2)$, with $P$ the output power, $\nu_0$ the laser frequency, $\nu_c$ the laser cavity linewidth; and also including the Henry enhancement factor $\alpha$ (typically between 2 to 7 for semiconductor lasers and lower for solid state lasers)~\cite{YarivYehBook}. This noise is especially high for ECDL and DBR due to the limited power that can be extracted from side-emitting diodes (below 100 mW) as well as by the large cavity bandwidth (GHz or more).
In contrast, a VECSEL would have orders of magnitude smaller Schawlow-Townes noise-floor thanks to the multi-Watt power accessible with the surface-emitting design as well as by the MHz linewidth of their cavity (associated to the much lower gain in this design). In fact, the measured noise level of the VECSEL at 10~Hz$^2$/Hz is probably limited by thermal and carrier-density fluctuation created by intensity noise of the pump, rather than by the Schawlow-Townes limit~\cite{Laurain2014,Hastie2023}. 
For the TiSapph laser, its noise level is so low that we cannot measure it with our finite measurement noise floor, in agreement with its expected Schawlow-Townes level at $10^{-5}$~Hz$^2$/Hz. 
The SIL laser display a noise level of 1~Hz$^2$/Hz, which is explained by the self-stabilization mechanism of the laser when coupled to a high-Q resonator~\cite{Hollberg1987,Maleki2015,Blumenthal2023}. Unfortunately, no fiber-based laser was available at the time of these measurements, but one could expect frequency noise at a level below 10~Hz$^2$/Hz~\cite{Dang:22}.

For all lasers, expect the VECSEL, we used home-made MZIs described previously and all measurements have in common a displayed noise level increasing below 100 kHz, originating from the MZI and not the lasers. In contrast, for the VECSEL, we used a commercial phase noise analyzer (OE4000 from OEwaves), whose performance remains excellent at lower Fourier frequencies. We note that, at MHz Fourier frequencies, the commercial system detection noise floor is 1~Hz$^2$/Hz, while the simple MZI can reach a 0.1~Hz$^2$/Hz level. We could even push down to 0.01~Hz$^2$/Hz (see the TiSapph curve in Fig.~\ref{fig:noisecollection}(a)), by using a single biased photo-detector and were then limited by photon shot noise. We also compare, in Fig.~\ref{fig:noisecollection}(e), the measured noise with our home-made MZI and with the commercial phase noise analyzer for the SIL laser. The agreement is very good, except in the 10 kHz range where the MZI exhibits an excess noise, described earlier. 

In Fig.~\ref{fig:noisecollection}(b), we compare the noise spectra of four lasers of identical architecture (ECDLs), from the same manufacturer but with different wavelengths, and obtain very similar frequency noise characteristics.
In Fig.~\ref{fig:noisecollection}(c), we compare the fast frequency noise of an ECDL laser before and after frequency doubling from 960 to 480 nm using SHG (second harmonic generation) in an enhancement cavity (with a linewidth of ~100 kHz): as expected, the noise power is multiplied by 4 to 10 at low Fourier frequency, but is filtered above the cut-off frequency of the cavity. This filtering action of a narrow linewidth cavity is used in Ref.~\cite{Levine2018} to passively reduce the noise of diode lasers. 
In Fig.~\ref{fig:noisecollection}(d), we illustrate the effect of a feedback loop on the noise of a DBR laser: for the fastest (PI corner at 500 kHz) and highest gain setting, the noise can be reduced below 300 kHz, but a servo bump appears at 1 MHz. In the next section, we will see how the fast feedforward noise eater is effective in this region where the feedback approach fails. 


\section{feedforward correction}
\label{sec:correction}

Having described how the MZI acts as a simple and fast frequency discriminator, we now shift our focus to reducing the phase/frequency fluctuation by applying a feedforward correction on a fast phase modulator~\cite{Hashemi2009,Hashemi2010,Lintz2017,Covey2022,Tey2023}.
As mentioned previously, while it is common to use a feedback loop to stabilize a system, this approach is plagued with finite loop time of several tens to hundreds of nanoseconds which strongly limits the possibility to reduce fast, sub-microsecond ($>$ MHz), fluctuations.
Here, a feedforward correction is more adapted as the delay in processing the signal (detection, amplification, filtering, ...) can be effectively canceled by delaying the optical signal by the same amount in several meters of fiber. 
On the other hand, this approach requires fine tuning to efficiently cancel the noise: the overall gain has to be set precisely to unity and the delay (phase) to zero. This can be compared to the more relaxed requirement of a feedback loop where one aim for the largest gain possible and a phase lag below 180$^\circ$. 

We will now explore the requirement on the feedforward path to reach large noise reduction, describe the experimental setup and discuss the results. 

\subsection{Theory}

As described in the previous section, the MZI frequency discriminator output is proportional to frequency fluctuation $\tilde{V} = s \tilde{\nu}$. As we apply the correction with a phase modulator, we need to convert this signal to be proportional to phase fluctuation $\tilde{\phi}$.  
We realize this step using a first-order passive low-pass filter (LP) with response:
\begin{equation}
s_{LP} = \frac{1}{1+i f/f_c}.
\end{equation}
Well above the cut-off frequency $f_c$, this filter effectively integrate the output of the MZI giving $\tilde{V} \approx -i s f_c/f \times \tilde{\nu} = - s f_c \tilde{\phi}$ which is proportional to phase fluctuation. 
We also introduce the phase modulator characteristics $V_\pi$ (the voltage for which a $\pi$-phase shift is applied), an overall variable gain $G$ and delay $\tau_d$, and note that the polarity of the correction is adjusted by locking the MZI interferometer either on the positive or negative slope. This leads to a Fourier-frequency-dependent correction $C(f) \tilde{\phi}$ applied by the phase modulator with:  
\begin{equation}
\label{eq:feedforward}
C(f) = \pm \frac{\pi f_c G s_0}{V_\pi} \frac{if/f_c}{1+if/f_c} \frac{\sin{(\pi f \tau)}}{\pi f \tau}e^{-i 2\pi f \tau_d}.
\end{equation}
The noise (in PSD) will be reduced as $S_\nu^{\rm out}/S_\nu^{\rm in} = |1-C|^2$, thus requiring to adjust the overall correction $C$ as close as possible to 1.





\paragraph{Precision on the adjustable settings}
Let's first consider how precisely the gain and delay have to be set to reach a noise cancellation of $|1-C|^2 = -30$ dB~\cite{Tey2024}. For a pure gain mismatch, this requires $|C|$ to be within 3 \% of unity, which can be obtained by adjusting the variable gain $G$. For a pure delay $|1-C| \approx \arg(C) = 2\pi f \tau_d$, the requirement on the phase error is $\arg(C) < 0.03$ rad (1.7$^\circ$), or, if expressed as a delay, $\tau_d < 0.5$~ns at $f = 10$ MHz which corresponds to a cable length of 10 cm.
\\
\paragraph{Frequency-dependence of the LP and MZI}

Two other sources of imperfection are the frequency-dependent response of the low-pass filter and the MZI discriminator, which are explicitly written in Eq.~(\ref{eq:feedforward}) and shown in Fig.~\ref{fig:feedforward_results}(c).

The MZI discriminator response includes a pure delay (absorbed in the definition of $\tau_d$) and a frequency-dependent gain error increasing as $C \simeq 1 - x^2/6$ (with $x = \pi f\tau$). Operating at $-30$ dB of noise reduction restricts the Fourier frequency to $f \lesssim 0.15 / \tau$, so below 7.5 MHz for the MZI delay $\tau = 20$~ns used later. 

The LP filter converges to a pure integrator only for $f\gg f_c$, with a LP gain error decreasing as $1/(2y^2)$ with $y = f/f_c$, and a phase error converging much slower as $1/y$. Consequently, $f>30 f_c$ is required to reach the 0.03 rad error target and thus the -30 dB cancellation level. Adjusting the overall delay, it is possible to cancel the phase error (and obtain an improved noise cancellation) at a finite Fourier-frequency by sacrificing response at larger $f$, as shown in Fig.~\ref{fig:feedforward_results}(c). Such peaked performance at the zero-crossing of the phase error can be seen in Ref.~\cite{Tey2024}.

The range of Fourier frequency $30 f_c <  f < 0.15/\tau$ of efficient feedforward operation can be widened by decreasing the cut-off frequency $f_c$ and the MZI delay $\tau$, however this comes at the cost of decreasing the MZI sensitivity  $s_0$ and more filtering of its output by the LP filter, such that a larger gain $G$ is required.


\subsection{Implementation}

Figure \ref{fig:feedforward scheme} shows the full setup. The laser light to be stabilized is first split in two paths: a measurement path with around 100~$\mu$W sent to a MZI discriminator, and a correction path consisting of a $\sim 10$~meter delay-line and a phase modulator. After correction, the light is sent to a second MZI discriminator to characterize the efficiency of frequency/phase noise reduction.

The MZI setup has been described in the first section. Its output now goes to a delay line consisting in a few meters of BNC cable set with a precision of 10 cm (0.5 ns, SRS DB64). We then adjust the feedforward gain with a combination of fixed attenuators with steps as small as 10~\% (1 dB), while for finer adjustment we attenuate the optical power to the MZI (Thorlabs V800PA). The signal then goes to a fixed-gain amplifier (iXblue, DR-VE-0.1-MO, 26 dB, $>$ 100 MHz bandwidth). The noise introduced by the amplifier is dominated by the input noise originating from the amplified photo-detector and can be neglected.  

The output of the amplifier is connected to an RC low-pass filter (R = 1 k$\Omega$, C = 2.7 nF, $f_c = 60$~kHz) realized with SMD components to reduce parasitic inductance and ensure correct operation up to 50 MHz (we observed a resonance at 100 MHz). The signal then finally goes to the fiber EOM (iXblue, NIR-MPX800-LN-0.1, $>$100 MHz bandwidth, $V_\pi = 1.1$~V), whose high input impedance of 10 k$\Omega$ insures that the filter is not loaded. 
We measured that the gain flatness of the correction system (amplifier, filter, EOM response) was better than the 3~\% required for a 30~dB noise cancellation over the 1-10 MHz range.


In contrast to a previous work using a free-space electro-optic modulator~\cite{Covey2022}, the use of a fiber EOM drastically decreases the requirement on the driving voltage, allowing to easily reach much larger bandwidth. However, fiber-based EOMs typically specify a maximum input power of 10-100 mW due to the photo-refractive effect, while free-space EOMs do not have such limitations.

\subsection{Results}

\begin{figure}
    \centering
    \includegraphics[width=\columnwidth]{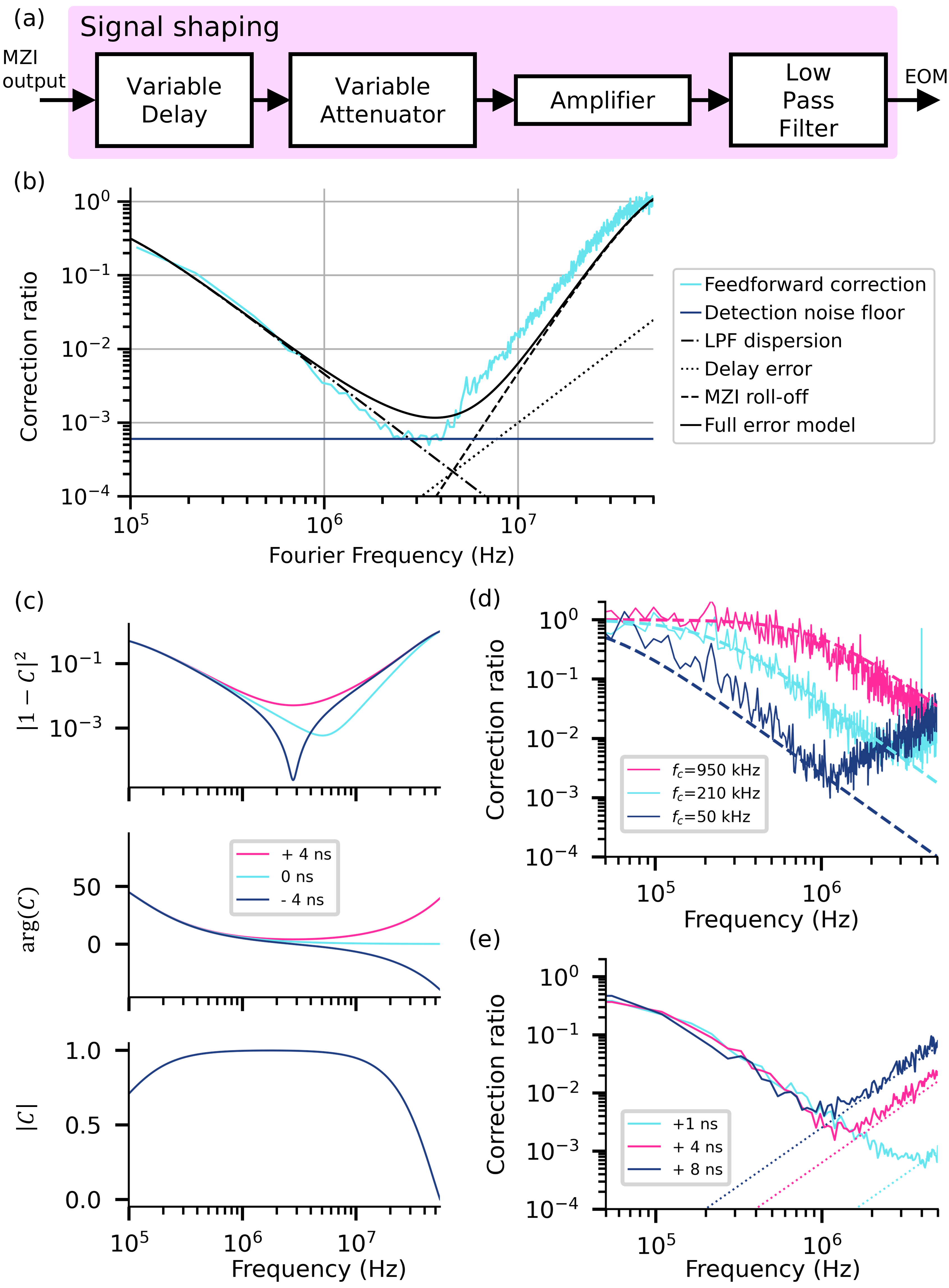}
    \caption{Feedforward correction results. (a) Elements in the feedforward path. (b) Noise correction (in ratio of PSD) achieved experimentally (light blue) compared to the estimated limits set by the noise floor (dark blue), LP filter dispersion (dot-dashed line), MZI discriminator roll-off (dashed), delay error (dotted) and the sum of these (solid black line). (c) Estimated noise reduction $|1-C|^2$ (in dB) as a function of Fourier-frequency $f$ (top) originating from the phase error $\arg{C}$ (middle) and gain error $|C|$ (bottom). Three different length of delay line are shown. (d) Noise reduction for several LP cut-off frequencies $f_c$ and model (dashed). (e) Noise reduction for several delays and model (dotted).}
    \label{fig:feedforward_results}
\end{figure}

The performance of the phase noise eater were briefly presented in Section~\ref{sec:overview}. Here, we provide a deeper analysis with a comparison to the expected limits. As shown in Fig.~\ref{fig:feedforward_results}(b), the measured correction (light blue line) is in excellent agreement with the model (solid dark line). 

We now go into each of the contributions to the full model. First, a frequency-independent limit of correction of 32 dB (dark blue line) is set by the photon shot noise and the electronic noise floor of the photo-detector that are both on the order $\rm 0.1~Hz^2/Hz$ (see Fig. \ref{fig:MZI_discriminator}). This limit is reached around $f = 3-5$~MHz. Second, at lower frequency, the correction is limited by the LP filter dispersion (Fourier-frequency-dependent phase-lag), as demonstrated by the perfect match with the dot-dashed line. We further illustrate this by varying the cut-off frequency ($f_c$) in Fig.~\ref{fig:feedforward_results}(d). Third, at the highest frequencies, the MZI response roll-off explains the degrading performance of the feedforward correction (dashed line). 
Fourth, we show with the dotted line the contribution of a 0.5 ns mismatch of delay (the smallest resolution we had), which is not expected to contribute at this level. For illustration, we also show, in Fig.~\ref{fig:feedforward_results}(e), how the correction is affected by larger delays of 4 and 8 nanoseconds set on purpose. Finally, by including all four independent imperfections into a single curve (solid line), we obtain a good agreement with the measured correction ratio.

Having discussed the improved phase stability in the 0.1-10 MHz range, we now consider if significant noise could have been added in the lower frequency range. One contribution could be the noise of the first MZI which is then fed to the fiber-EOM. As seen in Fig.~\ref{fig:noisecollection}(e), the MZI noise is at the $10^2$~Hz$^2$/Hz ($10^4$~Hz$^2$/Hz) level at $f=10$~kHz (1~kHz). If this noise would be imprinted directly to the laser, it would contribute a gate error of $3 \times 10^{-5}$~\cite{tsai2024}. 
However, we realize the integrator performing frequency-to-phase conversion before the phase-modulator (fiber-EOM) with a low-pass filter of cut-off frequency $50$~kHz, which is equivalent to an ideal integrator followed by a high-pass filter. Consequently, the MZI noise imprinted onto the laser is largely suppressed below 10~kHz, further reducing its contribution to gate error by a factor of more than 30, making this noise source negligible in the current noise budget~\cite{tsai2024}.
A second contribution could be the noise imprinted by the long fiber delay line (10~m), which is sensitive to thermo-mechanical noise. Previous studies have measured excess frequency noise on a level of $10^2$~Hz$^2$/Hz in presence of 80~dB of acoustic noise~\cite{Pang1992,Ma1994}. While such noise would be deleterious when operating at low Rabi frequencies (such as in metrology), for gate operations with MHz-range Rabi frequencies it is again negligible.

Following these observations, further improvements of the phase noise eater can be envisioned. The low and high frequency response can be improved by having a slower filter (decrease $f_c$) and faster MZI discriminator (decrease $\tau$), and by optimizing the amplifier to compensate for the decreased sensitivity . Also, an automatic gain control procedure acting on the continuously-variable optical attenuator, would allow to keep optimized conditions even for drifting laser power at the MZI input that otherwise disturbs the overall gain~\cite{Tey2024}.

Finally, we comment on the influence of the phase noise eater on intensity noise. Electro-optic modulators (either free-space or waveguide-based) used for applying the phase correction also slightly affect the intensity, an effect called Residual Amplitude Modulation (RAM). For our device, we measured a ratio of phase to intensity modulation of $k = \rm 0.1\%/rad$. We note that the RAM depends strongly on the DC bias applied to the EOM (probably caused by some etalon effect), and can be further decreased by a factor 30 for an optimal bias. Operating at such a sweet spot would however require a dedicated locking procedure of the bias point as it otherwise drifts. For this discussion, we picked the point of maximum RAM. The excess RIN (relative intensity noise) caused by the feedforward correction is simply ${\rm RIN} = k^2 \times S_{\phi}$. 
In practice, for the laser shown in Fig.~\ref{fig:feedforward scheme}, with a measured frequency noise of 1000~Hz$^2$/Hz and RIN of $-150$ dBc/Hz, we observed that the RIN doubles at $f= 1$~MHz (the excess RIN from the feedforward is equal to the laser RIN). We conclude by pointing out that such low level of RIN have a negligible contribution to the error budget of quantum gates~\cite{tsai2024}. 

\subsection{Discussion}

We now compare the performance described above to other approaches. First, compared to a feed-back stabilization, where noise suppression can easily reach 50 dB at kHz frequencies but fails at MHz due to the finite loop time and bandwidth limitations, the feedforward approach is very effective in cancelling noise between 1 to 10 MHz. With respect to filtering by a cavity where noise is suppressed as $(f_c/f)^2$ for $f\ll {\rm FSR}$ ($f_c$ is the cavity linewidth, FSR the free spectral range), we have similar noise reduction as with $f_c \sim 100$~kHz up to 3 MHz (where we reach peak performance), while for higher Fourier frequencies our feedforward method suffers from the limited bandwidth of the interferometer (50 MHz), much smaller than the typical bulk cavity FSR in the GHz range. Other differences are the more limited output power from a filtering cavity (typically sub-mW) compared to the phase noise eater limit of the electro-optic modulator (10-100 mW for a waveguide-version, Watt-level in free-space); as well as in the reduced footprint and complexity of the fully-fiberized phase noise eater. 

Finally, other feedforward noise suppression results have already been reported in the literature. With the same configuration as reported here, Hashemi and coworkers reduced the frequency noise of a DBR laser with a white noise floor at $S_\nu \sim 10^6 \, {\rm Hz}^2/{\rm Hz}$ by 15 dB for $f$ up to 100 MHz~\cite{Hashemi2009,Hashemi2010}. This large bandwidth was made possible by using a smaller delay in the interferometer at the cost of a reduced sensitivity (and thus increased noise floor), which is reasonable there given the large initial noise of the DBR laser used in their work. Other approaches than using a delayed interferometer for frequency noise measurement and feedforward are also possible. In Ref.~\cite{Covey2022}, the authors perform an heterodyne measurement between the noisy laser and its filtered version at the output of a cavity, and demonstrated 20 dB suppression at sub-MHz Fourier frequencies and 3 dB at 5 MHz. In Refs.~\cite{Tey2023,Tey2024}, Chao, Hua and coworkers instead directly use the same Pound-Drever-Hall error signal as used for feedback stabilization on a reference cavity, and feedforward its high-frequency components, directly proportional to phase noise, to an electro-optic modulator. They achieve 30 dB of cancellation in the $0.3-3$ MHz range, a broader range than with our MZI measurement thanks to the wider region where the PDH signal is proportional to phase noise (for $f_c \ll f \ll {\rm FSR}$). The absolute noise was however not reported (only the relative cancellation for an intentionally added modulation), neither the measurement noise floor, so it is yet unclear if the noise can be pushed below the 10~Hz$^2$/Hz level. Such cancellation performance were also later reproduced independently, achieving 30 dB at 2 MHz~\cite{Cornish2024}.


\section{Atom interrogation}
\label{sec:atom}

\begin{figure}
    \centering
    \includegraphics[width=\columnwidth]{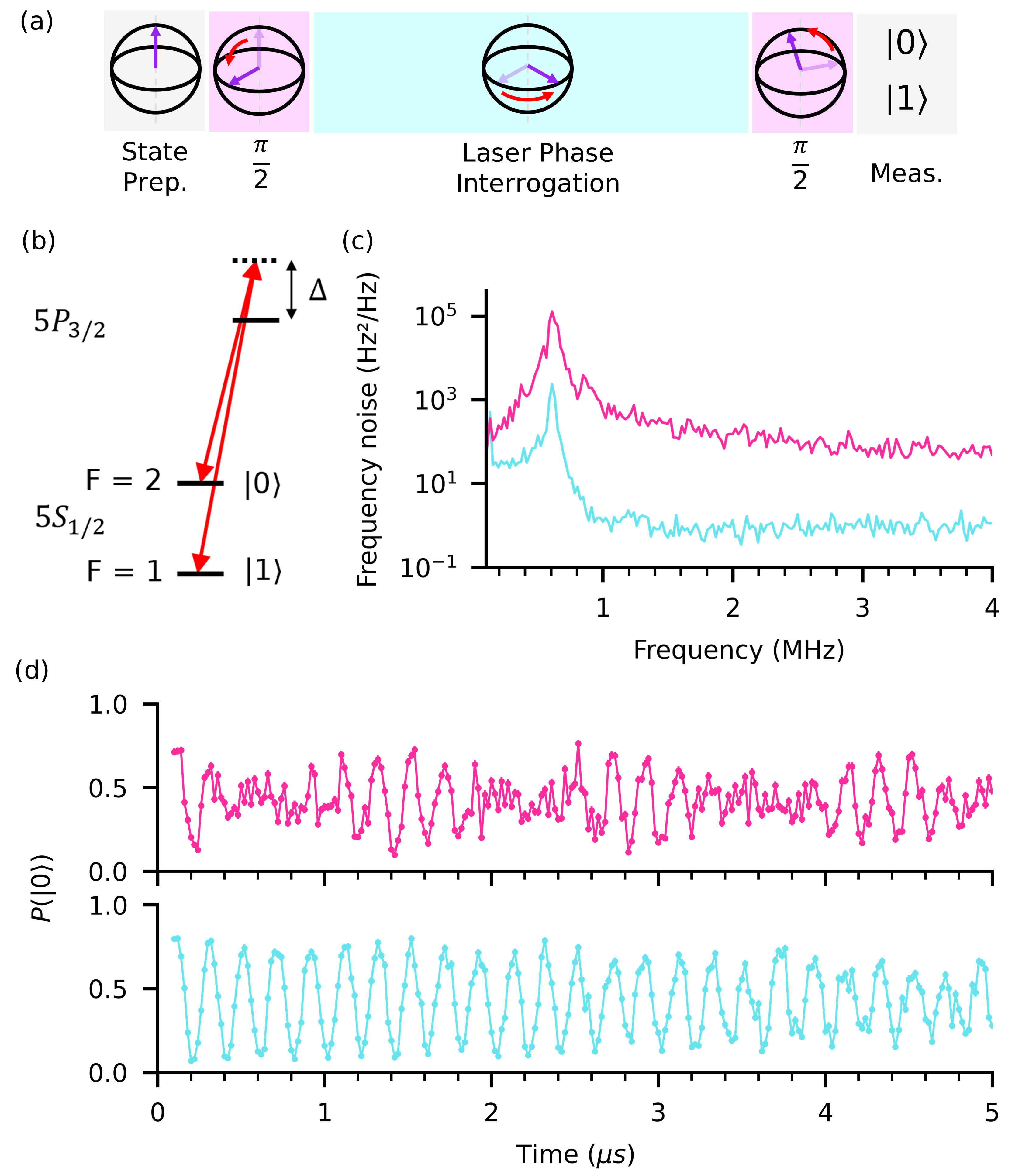}
    \caption{Atom interrogation. (a) Ramsey interferometry scheme for sensing the laser phase noise. 
    (b) Energy levels and the two laser drives involved in the Raman transition. 
    (c) Frequency noise spectrum of one of the laser drive with (without) phase noise cancellation in light blue (pink). The exaggerated servo-bump is reduced by more than 20~dB, as well as the white-noise level initially around 100 Hz$^2$/Hz. (d) Ramsey fringes on the clock states with (without) phase noise cancellation in light blue (pink), showing clear improvement with the phase noise eaters.}
    \label{fig:atom_interrogation}
\end{figure}

In the preceding sections, we have detailed our approach for measuring and quantifying phase noise within a laser, and presented a feedforward scheme to actively diminish this noise. In this fourth part, we investigate \textit{in situ} interrogation of cold $^{87}$Rb atoms and observe improved fidelity of manipulation thanks to the noise reduction. Such improvements after phase noise reduction have been observed in Rydberg excitation~\cite{Levine2018} and molecular state transfer~\cite{Cornish2024}. 



The experiment is performed on a $^{87}$Rb platform previously described~\cite{Chew_Ultrafast}, with an array of up to hundred atoms.  
We did not operate on the Rydberg transition, but rather on a Raman transition between the two ground states, see Fig.~\ref{fig:atom_interrogation}(b). It has the advantage of removing the Doppler and E-field noise present for Rydberg states, thus better isolating the effect of laser phase noise.
We interrogate the atoms with a Ramsey sequence, shown in Fig.~\ref{fig:atom_interrogation}(a), that can be interpreted as an atomic MZI with the atoms serving as a stable phase reference. 
We use two independent 780~nm ECDL laser sources, detuned by the intermediate  5P$_{3/2}$ state by 3 GHz, and frequency locked 6.8 GHz apart on independent frequency references: a high-finesse Fabry-Perot cavity for laser 1 and a spectroscopy cell for laser 2. Both lasers have their phase stabilized by two phase noise eater systems. The frequency noise PSD of laser 1 is displayed in Fig.~\ref{fig:atom_interrogation}(c). The frequency lock was badly tuned on purpose to enhance the servo-bump, strikingly visible at 700 kHz. When the phase noise eater is activated, the PSD is decreased by more than 20 dB.  

The measured Ramsey fringe is presented in Fig.~\ref{fig:atom_interrogation}(d) and is described by:
$$P_{\rm 0}(t) = \frac{1}{2} + \frac{1}{2} \cos \left[ (\delta t + \Delta\phi(t) \right] $$
with $\delta = 2\pi \times 4.8$~MHz the chosen two-photon detuning, and $\Delta\phi(t) = \phi(t) - \phi(0)$ the phase difference after an interrogation time $t$.
Without feedforward stabilization, we observe Ramsey fringes with little overall damping but a contrast periodically reaching near zero.
To interpret this, it is useful to look at the phase noise spectral density of the lasers. 
Since the PSD is strongly peaked around $f_p$, we can approximately model the laser phase as $\phi(t) = a\times \cos (2\pi f_p t + \theta)$, with $\theta$ a random phase with uniform sampling (for each experimental realization), and $a$ a random amplitude described by a Rayleigh distribution with rms given by the phase noise measurement~\cite{Saffman_gatefidelity_2022}. 
Consequently, the variance of $\Delta \phi$ periodically reaches a maximum (minimum) at odd (even) integer values of $2 f_p \tau$.
This model is in agreement with the results shown in Fig.~\ref{fig:atom_interrogation}(d). At $t = 1/(2 f_p) = 0.7~\mathrm{\mu s}$, the fringe's contrast is 0 and it revives fully at $t = 1/f_p = 1.4~\mathrm{\mu s}$. 
When we turn on the phase noise eater, the modulation of the Ramsey fringes completely disappears as the laser phase is sufficiently stabilized. 

We conclude this section by noting that the interrogation of atoms to judge the level of phase noise was not very sensitive. We had to increase the phase noise, by badly tuning the feedback loop, to see a clear effect on the Ramsey signal. More refined measurements could lead to a better sensitivity~\cite{tsai2024,Cornish2024}.

\section{Conclusion and Outlook}

We have described a simple, and powerful, technique to quantify the fast phase noise of laser sources. While this is far from a new technique, as it dates back from the invention of the laser~\cite{Armstrong1966,Sorin1992}, we believe that its importance for high-fidelity manipulation of the electronic states of neutral atoms, ions, and molecules, called for a modern presentation with all information needed to easily reproduce the measurement setup. 

We then characterized a collection of lasers, which displayed frequency noise varying by more than 8 orders of magnitude depending on their design. In particular, we illustrated that semiconductor lasers, with advanced designs such as VECSEL or SIL lasers, have a satisfyingly low level of noise. While they do not reach the performance of a solid-state laser, it should be more than enough for high-fidelity manipulation. We expect that this comparison will guide experimentalists in the choice of the most adequate lasers. 

In the second part of this work, we showed how the phase noise measurement can be used for a feedforward correction of the phase. We achieved up to 30 dB of cancellation at a Fourier frequency of 3 MHz, and more than 20 dB in the 1-10 MHz, bringing the laser phase noise below the Hz$^2$/Hz level. The performance of the system was discussed and matched well with calculations. We finally highlighted the effect of improving the laser phase noise when manipulating the coherence of an array of individual neutral atoms. 

We envision that this phase noise eater could be used with standard, low-cost, easy to assemble, semiconductors-based ECDLs whose phase noise is however unacceptable for the most demanding applications; instead of opting for the more expensive commercial products based on VECSEL and SIL. Combined with the usual PDH technique, it offers a low-noise laser source over the Fourier spectrum ranging from a few Hz to 10 MHz, with applications for the manipulation of Rydberg atoms~\cite{Levine2018, Saffman_gatefidelity_2022, Covey2022, Gross2022, tsai2024}, atomic interferometers~\cite{Nazarova2008,Chiarotti2022}, for coherent transfer between molecular states~\cite{Ni2021, Luo2021, Cornish2024}, or for trapped ion quantum gates~\cite{Akerman2015}.


\begin{acknowledgments}
We thank A. Nakai, M. Suzui and S. Tsuruta for technical supports in the making of the fiber interferometers, and T. Nakamura for helpful discussions on the fiber stretcher. SdL thanks T. Wang, S. Hollerith and J. Guo for discussions on the phase noise, and T. Jaunet-Lahary for a careful reading of the manuscript. This work was supported by MEXT Quantum Leap Flagship Program (MEXT Q-LEAP) JPMXS0118069021 and JST Moonshot R\&D Program Grant Number JPMJMS2269.
\end{acknowledgments}

\bibliographystyle{apsrev4-1}
\bibliography{main}

\end{document}